\documentclass[twocolumn, notitlepage]{article}
\usepackage{graphicx} 
\usepackage{amsmath} 
\usepackage{xcolor} 
\usepackage{subcaption}
\usepackage{soul} 
\usepackage{flafter}
\usepackage{float}
\usepackage{dblfloatfix}
\usepackage{authblk}
\usepackage{url}

\author[1]{Ville Jantunen}
\author[1]{Daniel R. Mason}
\author[1,2]{Sanjeet Kaur}

\affil[1]{UK Atomic Energy Authority, Culham Campus, Abingdon, Oxfordshire OX14 3DB, United Kingdom}
\affil[2]{Department of Engineering Science, University of Oxford, Parks Road, Oxford OX1 3PJ, UK}

\title{
Hydrogen Thermodiffusion in Tungsten: Trapped Hydrogen Distorts Heat of Transport Estimation from Total Concentration Profiles}
\date{\today}

\begin{document}


\twocolumn[
\begin{@twocolumnfalse}
\maketitle
\begin{abstract}
We investigate hydrogen diffusion in tungsten under thermal gradients in the presence of trapping sites using multiscale simulations. Using molecular dynamics simulations, we find that the heat of transport, which quantifies the coupling between temperature gradients and hydrogen transport, for hydrogen in tungsten is positive (thermophobic) at lower temperatures but changes sign near $1000$ K. We investigate the origin of the heat of transport and find that harmonic transition state theory with elastic thermal-expansion corrections is insufficient to explain the observed behaviour, consistent with either a breakdown of local thermodynamic equilibrium or non-negligible anharmonic contributions.

We further extend previous studies of hydrogen thermodiffusion by including the effect of trapping sites in both simulations and analytical models. We show that when trapped hydrogen accounts for a significant fraction of the total hydrogen concentration, the heat of transport inferred from measurements of the total concentration is dominated by the temperature dependence of the trapped hydrogen population and no longer reflects the gradient in the mobile hydrogen concentration. Consequently, experimentally measured heats of transport cannot in general be used directly in continuum-scale models that explicitly account for trapping and detrapping. In such models, the heat of transport should be based on mobile hydrogen concentrations only to avoid incorporating trapping effects twice.

\end{abstract}
\vspace{1em}
\end{@twocolumnfalse}
]

\section{Introduction}

Understanding hydrogen transport and retention in fusion materials is essential for the successful operation of deuterium–tritium reactors. These processes affect tritium inventory management, reactor safety, and the efficiency of the fuel cycle. During operation, hydrogen isotopes may be implanted into plasma-facing materials, migrate through the lattice, and become trapped at defects, resulting in substantial retention. Excessive retention decreases the amount of recoverable fuel, complicates tritium accounting, and contributes to the radioactive inventory of the reactor system.

Energy is deposited in fusion reactor materials through both plasma–surface interaction and neutron irradiation. In plasma-facing components, such as the ITER divertor, steady-state heat fluxes of $\sim$10 MW m$^{-2}$ and transient loads up to $\sim$20 MW m$^{-2}$ are expected, leading to near-surface temperature gradients on the order of 10$^2$ K mm$^{-1}$ \cite{ITER_divertor_2026, pitts2017physics}. These gradients influence the hydrogen diffusion through the Soret effect, which couples heat and mass transport. The strength of this coupling is quantified by the Soret coefficient or, alternatively, by the heat of transport.

In addition to thermal gradients, reactor materials are subjected to neutron irradiation, which generates defects that act as trapping sites for hydrogen isotopes \cite{causey2001use, roth2011hydrogen, hatano2013deuterium, tanabe2014review, mason2021parameter}, leading to increased retention and reducing permeation.

The transport of hydrogen in fusion materials has traditionally been described using diffusion–trapping frameworks by McNabb–Foster and Oriani \cite{mcnabb1963new, oriani1970diffusion}. These approaches have been widely implemented and applied to simulate hydrogen diffusion in fusion relevant settings \cite{hodille2015macroscopic, BENANNOUNE201942, dark2024modelling, delaporte20243d, dark2025festim, simon2025moose}. Recent developments have refined these descriptions by extending the classical frameworks to account for multi-occupancy traps \cite{schmid2014, hodille2016study, dark2025festim, kaur2025effect}, leading to more accurate representations of hydrogen permeation and retention. Nevertheless, the impact of such trapping processes on thermally driven transport remains less well understood.

Early experimental studies of hydrogen thermodiffusion, reviewed by Oriani \cite{ORIANI1969} and discussed by Weeks \cite{weeks1972}, reported both positive (thermophobic) and negative (thermophilic) heats of transport for hydrogen depending on the host material. Temperature-dependent heats of transport have also been observed more recently in fusion-relevant steels \cite{malo2025thermo}, with similar behaviour reported in earlier studies \cite{SUGISAKI198579}. Nevertheless, experimental data remains scarce, with tungsten notably absent from the available literature despite its importance in the fusion context. Estimates of the heat of transport of hydrogen in tungsten therefore rely on theoretical and simulation studies. In a recent contribution, Alturk et al. \cite{alturk2026thermal} used molecular dynamics simulations to estimate this quantity, reporting a negative Soret coefficient for H in W.

Molecular dynamics simulations provide important insight but are subject to inherent limitations. When diffusion barriers are high, low-temperature regimes become prohibitively expensive to study directly, and results must often be extrapolated from simulations performed at higher temperatures. Without an underlying theoretical model for thermally driven transport, such extrapolation may be unreliable. For example, Arrhenius rates appear nearly linear over a narrow high-temperature range, but linear extrapolation would lead to poor model of the exponential behaviour at low temperatures.  

In addition, classical molecular dynamics neglects electronic contributions, which can be important when electrons play a significant role in heat transfer and may influence transport through mechanisms such as electron wind forces \cite{ORIANI1969, SUGISAKI198579, long2026interplay, VonToussaint_discussion}. While two-temperature molecular dynamics can incorporate thermal conduction by electrons, it is not equipped to describe momentum transfer from electrons under thermal gradients and does not include forces arising from electric currents generated or changes in electric fields. Nevertheless, investigating the underlying non-electronic mechanisms remains valuable, although fully quantitative predictions should be complemented by analysis of electronic effects where relevant.

In this work, we investigate the combined effects of trapping and thermodiffusion using analytical models and multiscale simulations. We study the relationship between mobile and total hydrogen concentrations in the steady state reached under thermal gradients and discuss its importance for the interpretation of experimental measurements.
We revisit existing analytical descriptions of thermodiffusion and discuss how the heat of transport follows from the temperature dependence of hydrogen diffusivity. As this framework predicts only a positive heat of transport, in contrast to recent simulation studies, we investigate the sign and temperature dependence of the heat of transport using molecular dynamics simulations under a range of simulation conditions. 

\section{Theory and methods}

\subsection{Theory}

Thermodiffusion flux can be expressed as:
\begin{equation}
    J=-D(\nabla x + x\frac{Q\nabla T}{k_{\mathrm{B}}T^2}) \label{eq:thermodiffusion}
\end{equation}
where $x$ is the mobile concentration, and $Q$ is heat of transport which quantifies the concentration gradient response to temperature gradients. We take Eq. \ref{eq:thermodiffusion} as a definition for $Q$.

If $Q$ is dominated by change in diffusion hopping rate with temperature and assuming Arrhenius equation for diffusion, it can be shown to be equal to the diffusion barrier $Q=\Delta E$. The connection between $Q$ and the diffusion barrier has been discussed in the literature e.g. in Refs. \cite{ORIANI1969, SUGISAKI198579}.

More generally, Weeks et. al \cite{weeks1972} have derived an expression for $Q$ from the change of diffusivity:

\begin{align}
Q &= k_{\mathrm{B}}T^2 \left(\frac{1}{D}\frac{\partial D}{\partial T} -4W\right )\label{eq:Q_general_diffusivity}
\end{align}

 Where $W$ originates from Taylor expansion of probability of hopping in the direction of or against the direction of temperature gradient $p_\mathrm{left}=0.5 + W(T)\nabla T...$ and $p_\mathrm{right}=0.5 - W(T)\nabla T...$
 
 Neglecting the directional bias $W(T)$, $Q=\Delta E$ follows from Arrhenius form of diffusivity and for any linear function of temperature e.g. $D=T/\gamma$, $Q=k_{\mathrm{B}}T$. Notably any model for diffusivity in which diffusivity increases with temperature leads to positive $Q$. Negative $Q$ requires either diffusivity to decrease with temperature, or large enough directional asymmetry in transition rates, of which only the latter seems plausible. One candidate for rate asymmetry arises from linearly changing potential energy minima and saddle points, which leads to 
\begin{equation}
    Q=\Delta E + aT \label{eq:linear_in_T_Q}
\end{equation}
as shown in Appendix \ref{app:linear_in_T_Q}. If $a$ arises from the elastic energy change with $T$, it is negative for defects with positive relaxation volumes, such as interstitial H.

For constant temperature gradients, we can solve the steady state concentration profile $x(z)$ in 1D analytically for all the functional forms of $Q$ expressed here. From the definition of $Q$ in Eq. \ref{eq:thermodiffusion}, in steady state $J=0$:
\begin{align}
    0 &=-D(\nabla x + x\frac{Q\nabla T}{k_{\mathrm{B}}T^2}) \\
    -\frac{\nabla x}{x} &= Q\frac{\nabla T}{k_{\mathrm{B}}T^2} \\
    Q &= -\frac{\nabla x}{x\nabla T} k_{\mathrm{B}}T^2 \label{eq:Q_measure}
\end{align}

After change of variable from z to T makes the solution to constant $Q=\Delta E$ we get:

\begin{equation}
    x(T) = x_0\exp\left(\frac{\Delta E}{k_{\mathrm{B}}T(z)}\right) \label{eq:simple_solution}
\end{equation}
Where $x_0$ is obtained by normalizing to correct total hydrogen content. This is the solution when diffusion follows the Arrhenius equation. 

With any linear diffusivity with temperature, $D=cT$, Q=$k_\mathrm{B}T$, and the analytical solution is:
\begin{equation}
    x(T) = \frac{x_0}{T}
\end{equation}

And for $Q=\Delta E + aT$ from a tilted potential energy landscape, or other linear in T models for $Q$ e.g. one in Ref.\cite{LONGHURST198561}, we have
\begin{equation}
    x(T) = c_{\mathrm{norm}}\exp\left(\frac{\Delta E -aT\log(T/T_0)}{k_{\mathrm{B}}T}\right)
\end{equation}\label{eq:non_symmetric_solution}
$T_0$ can be chosen arbitrarily (e.g. 1 K) or absorbed into $c_{\mathrm{norm}}$. 

When multi-occupancy traps, with up to n H atoms bound to as single trap, are included, we can write the total hydrogen concentration as a sum of mobile and trapped concentrations. For a single trap species:
\begin{align}
    \label{eqn:total_conc}
    c &= x +c_\mathrm{trapped} \\
     &= x + \rho\langle s \rangle
\end{align}
where $\rho$ is the trap density, $\langle s \rangle=\sum_{i=0}^n iy_i$ is the average occupancy, which is a sum of i-occupied states times their probabilities $y_i$. $y_i$ is a function of the thermodynamic environment of the trap i.e. mobile concentration of H and temperature. This description can be trivially extended for multiple trap species by replacing $\rho\langle s \rangle$ with a summation over all different trap species.

We use the $y_i$ calculated by finding the steady state solution of trapping and detrapping rates as expressed in Ref. \cite{kaur2025effect} for tetrahedral occupation in bcc W, using binding energies from Ref. \cite{mason2023empirical}. 

\begin{align}
    n_H &= \int c(r)d^3r = A\int_0^L c(z)dz \\
    n_H/A & = \int_0^L x(z)dz + \int_0^L \rho\langle s(x(z),T) \rangle dz
\end{align}

As the mobile concentration profiles $x(z)$ are solutions to first order ordinary differential equations, they all contain one free parameter. We integrate this expression numerically and solve the free parameter iteratively by fixing $n_H/A$. For constant $Q$:
\begin{align}
        n_H/A = ~&x_0\int_0^L \exp \left(\frac{Q}{k_\mathrm{B}T(z)} \right) dz \nonumber \\ &+ \int_0^L \rho\langle s(x(z),T(z)) \rangle dz
\end{align}
where $x(z)$ also appears implicitly in $\langle s(x(z),T(z) \rangle$.

In order to express the relation between the temperature gradient and the total concentration we define $Q_\mathrm{eff}$ 
\begin{align}
    \label{eqn:eff-Q-definition}
    Q_\mathrm{eff} &= -\frac{\nabla c}{c\nabla T}k_\mathrm{B}T^2 \\
    &= -\frac{\frac{dc}{dT}}{c}k_\mathrm{B}T^2
\end{align}
Where the temperature derivative of the concentration depends on the mean occupation $\langle s(x(T),T) \rangle$, which itself is a function of $x(T)$ through trapping rates scaling with $x(T)$.
\begin{align}
    \label{eqn:total_c_derivative}
    \frac{dc}{dT} =\frac{\partial x}{\partial T} + \rho \frac{d}{d T}\langle s(x(T),T) \rangle
\end{align}
for this simplest form for $Q$, $Q=$~const., we then get an expression 
\begin{align}
    \label{eqn:eff-Q}
    Q_\mathrm{eff} = \frac{1}{x+\rho \langle s \rangle}\left[Q \bigl(x+\rho\mathrm{Var}(s)\bigr) + \rho\mathrm{Covar}(s,\epsilon) \right]
\end{align}
where $\epsilon_i=\sum_{s=1}^i E^b_s$ is the cumulative binding energy. Derivation of this expression, and discussion of it assumptions and physical interpretation, can be found in Appendix. \ref{app:qeff_derivation}.
\subsection{Simulation Methods}

We have used lattice Kinetic Monte Carlo (KMC) and molecular dynamics (MD) to simulate thermodiffusion of H in W. Our lattice KMC model is described in Ref. \cite{mason2026multiscale}. In summary, it consists of hydrogen in tetrahedral interstitial sites and vacancies. Binding energies of H in vacancies and diffusion barriers are parametrized to Mason et. al. 2023 \cite{mason2023empirical} potential. Our KMC model calculates transition rates based on a local temperature and includes a one dimensional constant temperature gradient, i.e. $T$ is only a function of $z$-coordinate, between the center the box and the z-edges. We use KMC for studying the trapping effects because trapping slows down the diffusion of hydrogen atoms and makes it difficult to reach steady state concentration profiles with MD. 

If $Q$ is split to three components, using the notation of \cite{SUGISAKI198579}, $Q=Q_{\mathrm{el}}+Q_{\mathrm{ph}}+Q_{\mathrm{int}}$, analytical models presented here and KMC address the intrinsic $Q_{\mathrm{int}}$ arising from changes in hopping rates with temperature. MD also includes $Q_{\mathrm{ph}}$ contributions. This decomposition is introduced solely to indicate which contributions to the heat of transport are captured by each modeling approach. It should not be taken to imply that the total heat of transport can necessarily be represented as the simple sum of independent contributions.

\begin{figure}[H]
    \centering
    \includegraphics[width=0.9\linewidth]{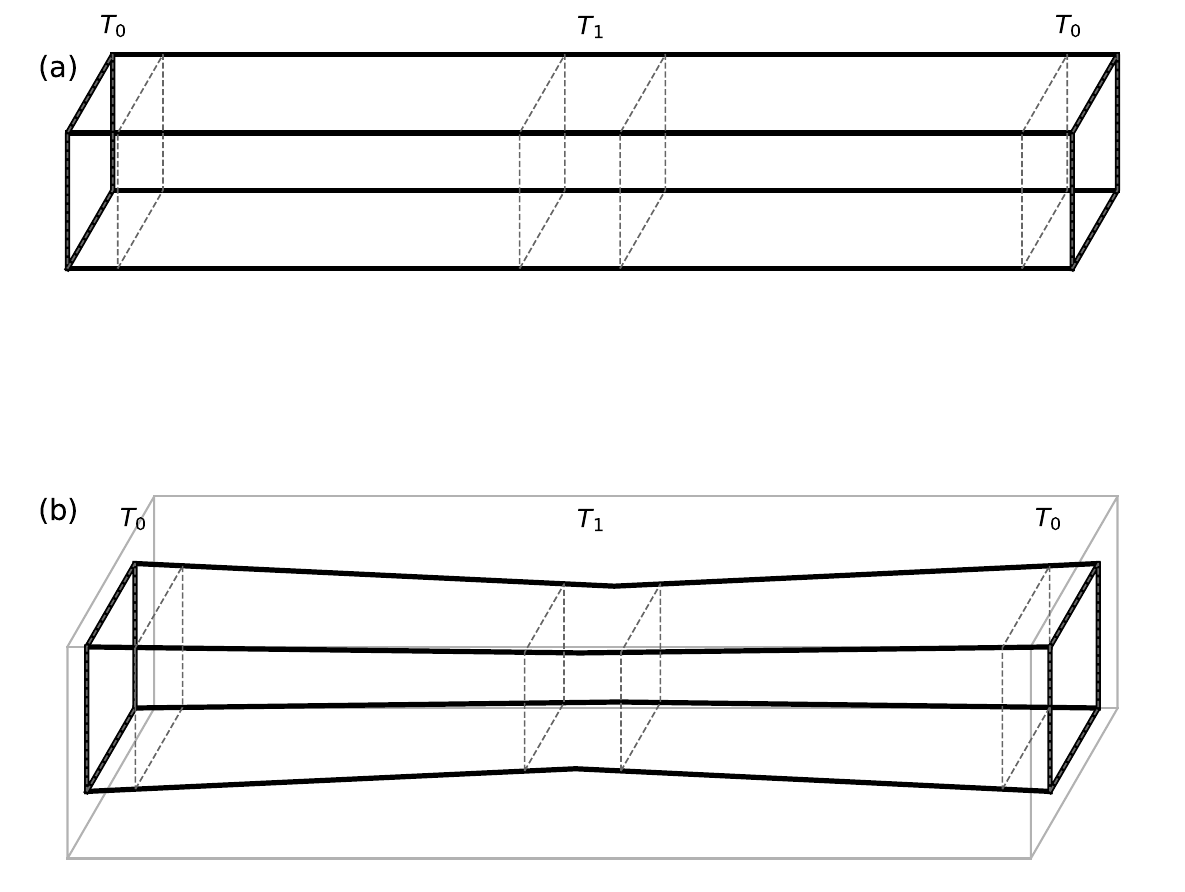}
    \caption{MD simulation geometry. $T_0$ and $T_1$ indicate the hot and cold thermostat regions, respectively. (a) Default setup with periodic boundary conditions in all directions. (b) Setup with open surfaces in the x and y directions to permit thermal expansion in all three directions. The outer box indicates the simulation cell, and the inner box represents the W rod. Thermal expansion is exaggerated for visual clarity.}
    \label{fig:simulation_setup}
\end{figure}

We performed our molecular dynamics simulations with LAMMPS \cite{LAMMPS} using multiple different setups. The primary configuration follows that of Ref. \cite{alturk2026thermal} extended to systems containing more than one hydrogen atom, shown in Fig. \ref{fig:simulation_setup}(a). An average hydrogen concentration of 0.5 at.\% was used, which is assumed to be sufficiently low to avoid significant concentration effects. In some cases where local concentrations became high, additional simulations at 0.05 at.\% were performed for comparison. All interactions in the system were described using the potential of Mason et al. (2023) \cite{mason2023empirical}.

Molecular dynamics simulations were performed using a periodic simulation cell consisting of $40 \times 40 \times 200$ W unit cells. In systems containing vacancies, 0.4~at.\% of W atoms were removed from the lattice to create vacancies. H atoms were then introduced randomly at an average concentration of 0.5~at.\%.

The system was equilibrated at 1500~K in NPT. After equilibration, a temperature gradient was established by introducing two thermostatted regions of width 10 unit cells. To check for thermostat effects, some systems were run with varying thermostat region thicknesses and Langevin thermostat damping parameters from 0.1 ps to 1 ps. Where not specifically mentioned, 0.1 ps is used. A constant simulation timestep of 0.2 fs was used in all simulations.

One thermostat was positioned at the centre of the simulation cell, while the second was applied at the boundaries in the $z$ direction. A Langevin thermostat was used to maintain the temperatures at 1000~K in the centre and 2000~K at the boundaries, and the rest of the system was simulated in NVT. Each simulation was run for approximately 20~ns to allow the system to approach steady state.

The use of a Langevin thermostat has been shown to reduce diffusion rates \cite{shiraishi2026}. In the present simulations, only W atoms are thermostatted, such that H does not experience direct frictional forces. Nevertheless, thermostat-induced artefacts are possible in the vicinity of the thermostatted regions.

Vacancies can in principle be introduced randomly; however, in this work a homogeneous distribution on a regular grid is employed. This choice avoids vacancy concentration fluctuations, and guarantees that no vacancies start near each other in order to reduce clustering. 

In the presence of vacancies, simulation times of 20~ns are insufficient to reach a steady state. However, we do not aim to present quantitative predictions for $Q_\mathrm{eff}$ from MD, and the obtained profiles provide a clear qualitative comparison between systems with and without trapping sites. We complement this simulation with another 20~ns simulation run where during the first 10~ns, thermostat temperatures were gradually decreased from +1000 K to their final temperatures. 

It should be noted that the imposed periodic boundary conditions introduce additional limitations. A free-standing tungsten sample subjected to a thermal gradient would expand non-uniformly, with greater expansion at the hotter regions. In a periodic simulation cell, which must remain parallelepiped, this variation in width is suppressed. As a result, thermal expansion perpendicular to the temperature gradient is artificially constrained, which may introduce non-physical stresses in the system. The resulting strain profile is a combination of thermal expansion and the Poisson effect, which causes increased local elongation in the hotter ends from the compressive stress and local contraction in the colder regions from the tensile stress. 

An alternative setup is to make the thermal gradient direction periodic and allow for open surfaces on other four faces, shown in Fig. \ref{fig:simulation_setup}(b). The challenge with this setup is that H immediately migrates to the surface of the box and diffuses on the surface, not in the bulk. We use LAMMPS fix wall/reflect to prevent this from happening. It should be noted that the solution is not ideal as such methods are not momentum conserving and the volume accessible for H atoms is constrained to a fixed parallelepiped region which does not change with thermal expansion. It is difficult to assess the importance of simulation artifacts resulting from the use of this method. In this setup the barostat is only applied in the $z$-direction and the temperature gradient is present from the beginning. After 40000 simulation steps we fix the outermost atoms in x-y directions to prevent the system from rotating and to continue the simulation in NVE + thermostatted regions. This setup is  compared to the default setup to analyse the magnitude of the effect from artifical strain in the parallelepiped periodic setup.

\subsection{Data Analysis}
We use histograms (with linear interpolation between bin centres) and Gaussian kernel density estimators (KDE) to represent concentration profiles. KDE yields smoother profiles; however, it tends to oversmooth the data near the edges and centre of the simulation box, where derivatives change abruptly. Where KDE is used in place of a histogram, this is explicitly stated. For the KDE, distances between evaluation points and data points are computed using the shortest distance consistent with periodic boundary conditions.

To characterize fluctuations in the steady-state profiles, the time series of concentration profiles was divided into temporal segments of equal duration. An average concentration profile was computed for each segment. The shaded bands represent the point-wise standard deviation across these segment-averaged profiles, evaluated independently at each temperature (position) coordinate. These bands provide an indication of  uncertainly in the data. However, no analysis of statistical independence between segments or correlations between spatial points is performed, and hence these bands should not be interpreted as rigorous error estimates. When evaluating $Q$ and $Q_\mathrm{eff}$ from concentrations profiles we use KDE when the data does not have clear wavelength dependent components. In the case of vacancies in a regular grid, the concentration profiles have a clear wavelength and we use truncated inverse discrete Fourier transform as a low-pass filter. Because this high-frequency component reflects genuine signal rather than noise, methods such as the Akaike information criterion do not provide a meaningful truncation point, as additional components continue to contribute information. Furthermore, a large fraction of the power spectrum resides in these high-frequency modes, meaning that retaining a fixed fraction of the spectrum is also not an appropriate criterion. We have verified that the chosen truncation point produces concentration profiles very similar to those obtained by averaging over the five nearest datapoints, while yielding a smoother profile. Although the exact truncation point affects the degree of oscillation in the reconstructed profile, it does not alter the overall trends. Since our primary interest is in these trends, we consider the present approach sufficient for the purposes of this work. While the exact values of $Q_\mathrm{eff}$ as a function of temperature are sensitive to the truncation choice, the resulting trends are a faithful representation of the underlying MD data.

\section{Results and discussion}\label{section:results_and_discussion}


\subsection{KMC in Absence of Traps}
\begin{figure}[H]
    \centering
    \includegraphics[width=1\linewidth]{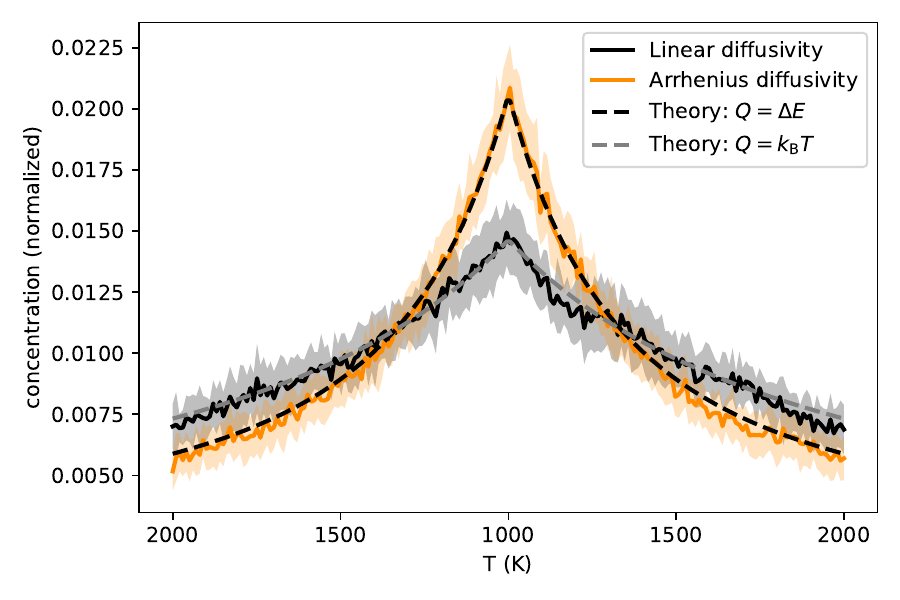}
    \caption{Hydrogen concentration profiles from KMC with Arrhenius (orange solid) and linear-in-temperature (black solid) diffusivity models, compared with their analytical solutions (grey and black dashed curves, respectively). Shaded bands indicate the pointwise standard deviation of profiles averaged over independent temporal segments of the trajectory. The system is symmetric about the center plane; unless otherwise stated, only  half of the simulation cell is shown in subsequent figures.}
    \label{fig:kmc_no_traps}
\end{figure}

Without any trapping sites present the theoretical prediction, $Q=k_{\mathrm{B}}T^2\frac{1}{D}\frac{\partial D}{\partial T}$ holds in the present KMC model for both Arrhenius and linear with Temperature functional forms of $D$ as shown in Fig. \ref{fig:kmc_no_traps}. 

\subsection{MD in Absence of Traps}
\subsubsection{Periodic Boundary Conditions vs. a Free-Standing Rod}

\begin{figure}[H]
    \centering
    \includegraphics[width=1\linewidth]{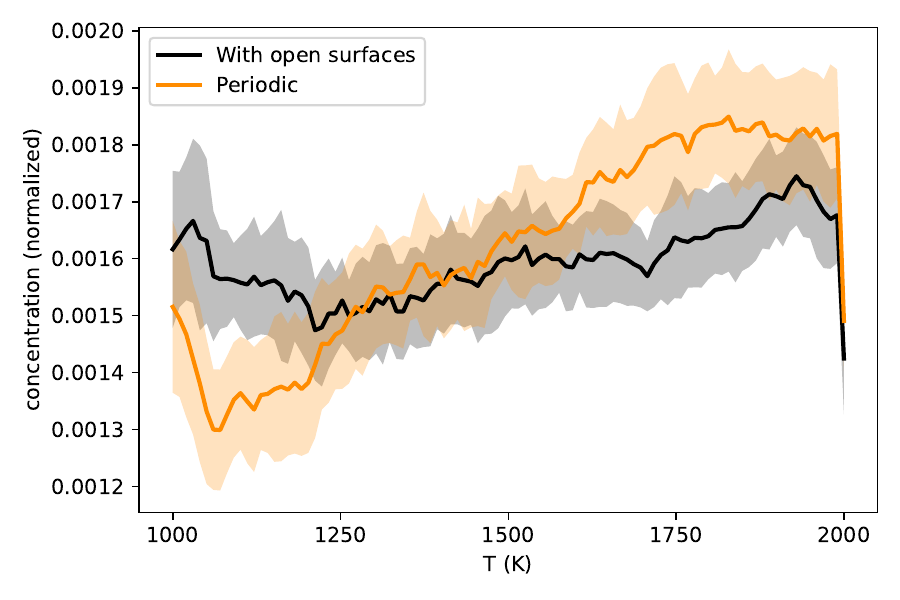}
    \caption{Normalised hydrogen concentration as a function of local temperature (1000--2000~K) from MD. Temperature is inferred from position assuming a linear gradient between the two thermostatted boundaries (1000 K and 2000 K). Black and orange curves correspond to a fully periodic cuboid equilibrated at 1500~K and a free-standing slab with open surfaces normal to $x$ and $y$, periodic in $z$ (thermal-gradient direction), respectively. Shaded bands show point-wise standard deviation between segment-averaged profiles.}
    \label{fig:md_orig_vs_open_surfaces}
\end{figure}

Because the KMC model includes only a limited subset of the relevant physics, we complement it with non-equilibrium molecular dynamics (NEMD) simulations, which naturally include additional physical effects such as elastic interactions, anharmonicity, and phonons. In the absence of vacancies (Fig.~\ref{fig:md_orig_vs_open_surfaces}), we observe that in the temperature range 1000--2000~K the heat of transport $Q$ is close to zero and slightly negative, corresponding to weakly thermophilic behaviour of hydrogen, qualitatively consistent with previous MD studies. The results from simulations with open surfaces and periodic boundary conditions are broadly consistent, both indicating that hydrogen migrates towards the hotter region, even though the curves do not overlap exactly. As discussed in the Methods section, the strain fields are not identical in the two simulation geometries, which may contribute to these small deviations.

\subsubsection{Lower Temperature Range (500-1500 K) and the Sign Flip of Q}

\begin{figure}[H]
    \centering
    \includegraphics[width=1\linewidth]{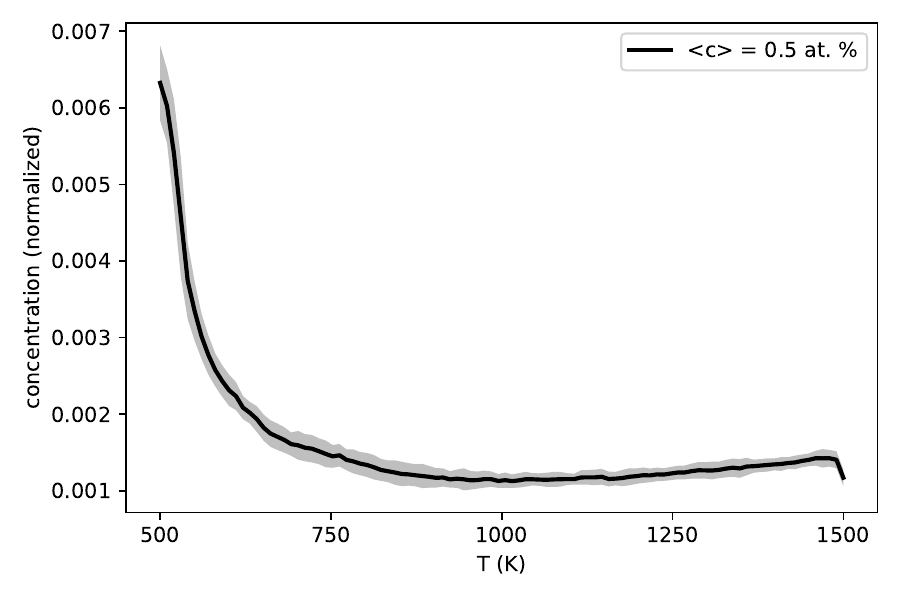}
    \caption{Normalised hydrogen concentration as a function of local temperature (500--1500~K, lower than in other simulations) from MD. Average Hydrogen concentration is 0.5 at.\%. Simulation was run for 80 ns. Shaded band shows point-wise standard deviation between segment-averaged profiles.}
    \label{fig:md_colder}
\end{figure}

From close inspection of Fig. \ref{fig:md_orig_vs_open_surfaces}, we notice that there might be a sign flip near the center (colder) thermostat, though from this data it is not possible to say if it is a thermostat effect or not. Hence we run another simulations from 500 K to 1500 K and from Fig. \ref{fig:md_colder} see that there is a sign flip point at $\sim 1000K$ from positive to negative $Q$. To confirm that high concentration near 500 K thermostat region is not an artifact, we have run different thermostat settings, of which 6 unit cell wide thermostat slabs with 0.5 ps Langevin thermostat damping were used in Fig. \ref{fig:md_colder}. Same conclusion are drawn from the simulations with stronger thermostatting and for 0.05 at. \% average concentration runs.

Linear in temperature $Q$ fits the data in Fig. \ref{fig:md_colder} well, and it might be tempting to attribute the linear behavior to elastic energy change from thermal expansion. Linear in temperature $Q$ from Linear in temperature change in potential energy landscape is described in Appendix \ref{app:linear_in_T_Q}. While a sign flip is qualitatively expected due to the barrier asymmetry that follows from potential energy gradient, the elastic energy change in this model is too weak to explain the effect. From linear elasticity theory, the elastic energy change with strain is $\Delta E = -P_{ij}\varepsilon_{ij}$. Here, the source of the strain is thermal expansion. After direct measurement of the strain tensor from MD data in the periodic system, this gives $\sim -2.4\cdot10^{-2}$eV change over 1000 K which means $a\sim -2.4\cdot10^{-5}$ eV/K. 

A direct fit to the concentration profile in Fig.~\ref{fig:md_colder} gives $\Delta E = 0.361 \pm 0.004~\mathrm{eV}$ and so $a = (-3.49 \pm 0.05)\times10^{-4}~\mathrm{eV\,K^{-1}}$. The reported errors are derived from the covariance matrix of the fit and provide an indication of the uncertainty in the fitted parameters. The difference between linear fit to $Q$ and a simple elastic energy correction model is an order of magnitude, so it is unlikely that sloped potential energy surface from thermal expansion alone would explain the observed behavior. Fitted $\Delta E$ also does not correspond to the migration barrier of the potential, which is $0.216$ eV. 

\subsubsection{Mass Effects on Q: 1 u vs. 10 u and the Applicability of HTST}
\begin{figure}[H]
    \centering
    \includegraphics[width=1\linewidth]{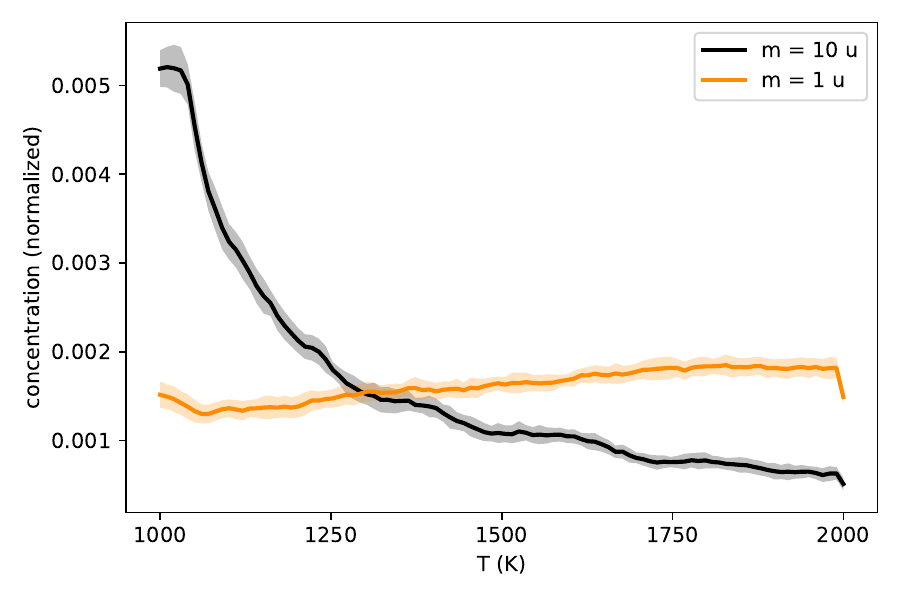}
    \caption{Normalised hydrogen concentration as a function of local temperature (1000--2000~K) from MD for hydrogen with an artificial mass of 10~u (black) and regular hydrogen (mass $\sim 1\,\mathrm{u}$) (orange). Shaded bands show point-wise standard deviation between segment-averaged profiles.}
    \label{fig:md_mass_comparison}
\end{figure}

\begin{figure}[H]
    \centering
    \includegraphics[width=1\linewidth]{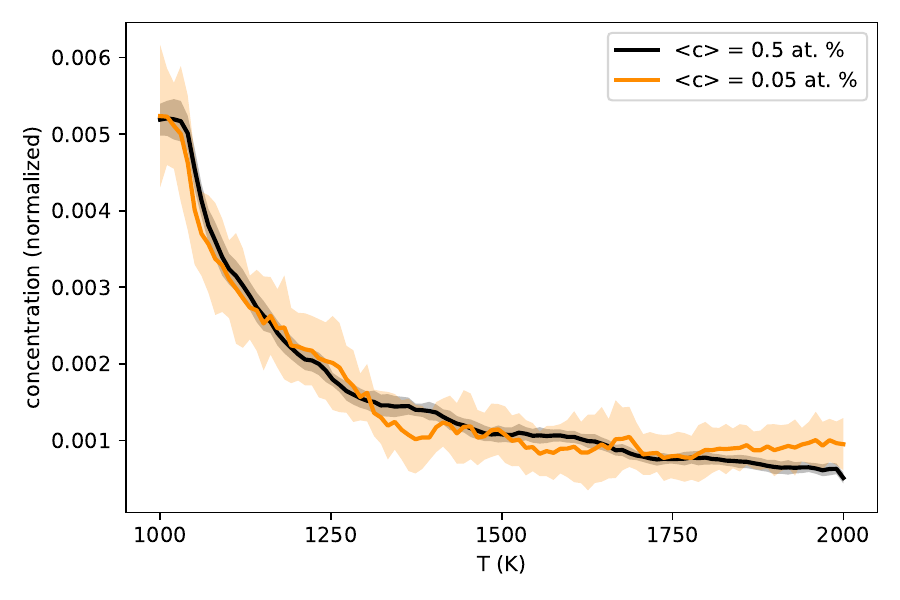}
    \caption{Normalised hydrogen concentration as a function of local temperature (1000--2000~K) from MD for H with artificial 10 u mass. Black and orange curves correspond to average concentrations of 0.5 and 0.05~at.\%, respectively. Shaded bands show point-wise standard deviation between segment-averaged profiles.}
    \label{fig:md_mass10_concentrations}
\end{figure}

To further investigate $Q$, we performed simulations with an artificially increased hydrogen mass of 10 u. The rationale is that, within classical MD, the potential energy surface is independent of atomic mass, as zero-point energy effects are absent.
Our theoretical predictions, based on biased diffusion barriers, assume that jump rates can be described using harmonic transition state theory (HTST) leading to an Arrhenius form. Within HTST, the atomic mass enters only through the prefactor, which is expected to be close to temperature-independent. While thermal expansion could introduce a temperature dependence in the prefactor through modifications of the potential energy surface, we have not observed evidence for such an effect, either in deviations from Arrhenius behaviour of the diffusion coefficient or in the temperature dependence of the vibrational (Einstein) frequencies of H.
A comparison between hydrogen masses of 1 u and 10 u is shown in Fig. \ref{fig:md_mass_comparison}. For hydrogen with mass 1 u, hydrogen is slightly thermophilic, whereas 10 u hydrogen is clearly thermophobic. This further indicates that elastic energy effects associated with thermal expansion are likely weak, since such effects would influence both mass cases similarly apart from minor changes in the prefactor scaling with temperature; absolute value of the prefactor is unimportant for determining heat of transport from steady state. To rule out concentration effects, we run an additional 10 u hydrogen simulation with 0.05~at.\% hydrogen concentration shown in Fig. \ref{fig:md_mass10_concentrations} and see no meaningful concentration effects.

These results suggest that the MD observations cannot be fully explained within harmonic or quasi-harmonic transition state theory (i.e.\ including thermal expansion effects alone), and instead point to the importance of additional anharmonic contributions. It is also possible that local thermodynamic equilibrium, i.e. the assumption that the local distribution function follows a Maxwell-Boltzmann distribution, breaks down. Such deviations would, in turn, affect the validity of harmonic transition state theory (HTST). Further discussion is presented in Appendix~\ref{app:break_of_local_eq}.

\subsection{Effects of Trapping Sites}

\subsubsection{Effects on Steady-State Concentration Profiles and $\mathrm{Q}_\mathrm{eff}$ Inferred from Experiments}
If we include vacancies in our KMC simulations, we get different concentrations profiles, depending on the concentration of traps. These profiles are shown in Fig. \ref{fig:sanjeet_vs_kmc}. We can see, that increasing number of vacancies biases the concentration profiles more towards the colder region of the box, even though $Q$ is the same in all simulations, as nothing has changed in the interstitial jump rates. This leads to an important implication: if $Q$ is measured from a steady state concentration in experiments, it is unlikely that the measurements allow for distinguishing between the hydrogen atoms in interstitial sites and within traps and hence the analysis is done on total hydrogen concentration. Measurements based directly on permeation flux should, in principle, yield the correct mobile-hydrogen heat of transport. However, the measured flux generally reflects the coupled effects of thermodiffusion, trapping, and detrapping kinetics, so the resulting $Q_\mathrm{eff}$ may not be readily separable from trapping-related contributions. Indeed, even under isothermal conditions the observed diffusion flux already contains contributions arising from trapping and detrapping processes, when trapping sites are present \cite{kaur2025effect}.

\subsubsection{Analysis of $\mathrm{Q}_\mathrm{eff}$ Based on the Analytical Solution}

\begin{figure}[H]
    \centering
    \includegraphics[width=1\linewidth]{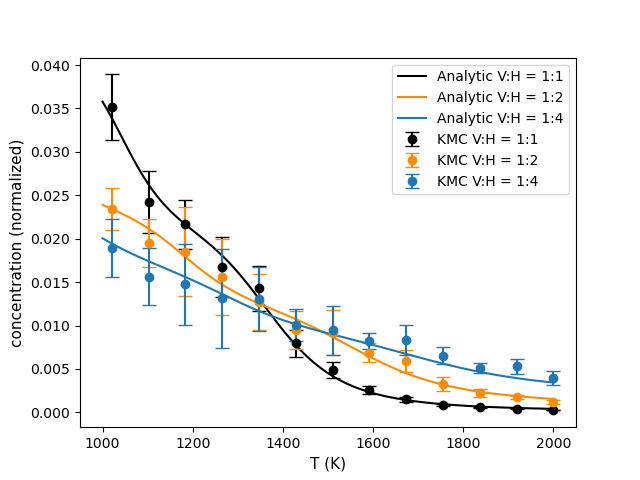}
    \caption{Normalized total hydrogen concentration as a function of local temperature. Black, orange, and blue curves correspond to systems containing 0.5, 0.25, 0.125 at.\% vacancies from local steady state analytic solution where mobile concentration profile is normalized iteratively. Markers correspond to KMC results. Vacancies are randomly spaced in the KMC simulations. Error bars show point-wise standard deviation between 10 independent KMC runs.}
    \label{fig:sanjeet_vs_kmc}
\end{figure}

The concentration profiles based on the analytical solution very closely follow those of KMC as seen in Fig. \ref{fig:sanjeet_vs_kmc}, so must $Q_\mathrm{eff}$. This allows us to analyse $Q_\mathrm{eff}$ directly using Eq. \ref{eqn:eff-Q}. Direct measurement of $Q_\mathrm{eff}$ from KMC and MD concentration curves is limited in the extent of the parameter space that can be explored within resonable computational resource use, and is discussed later in this section.


\begin{figure*}[!t]
    \centering
    \includegraphics[width=1.0\linewidth]{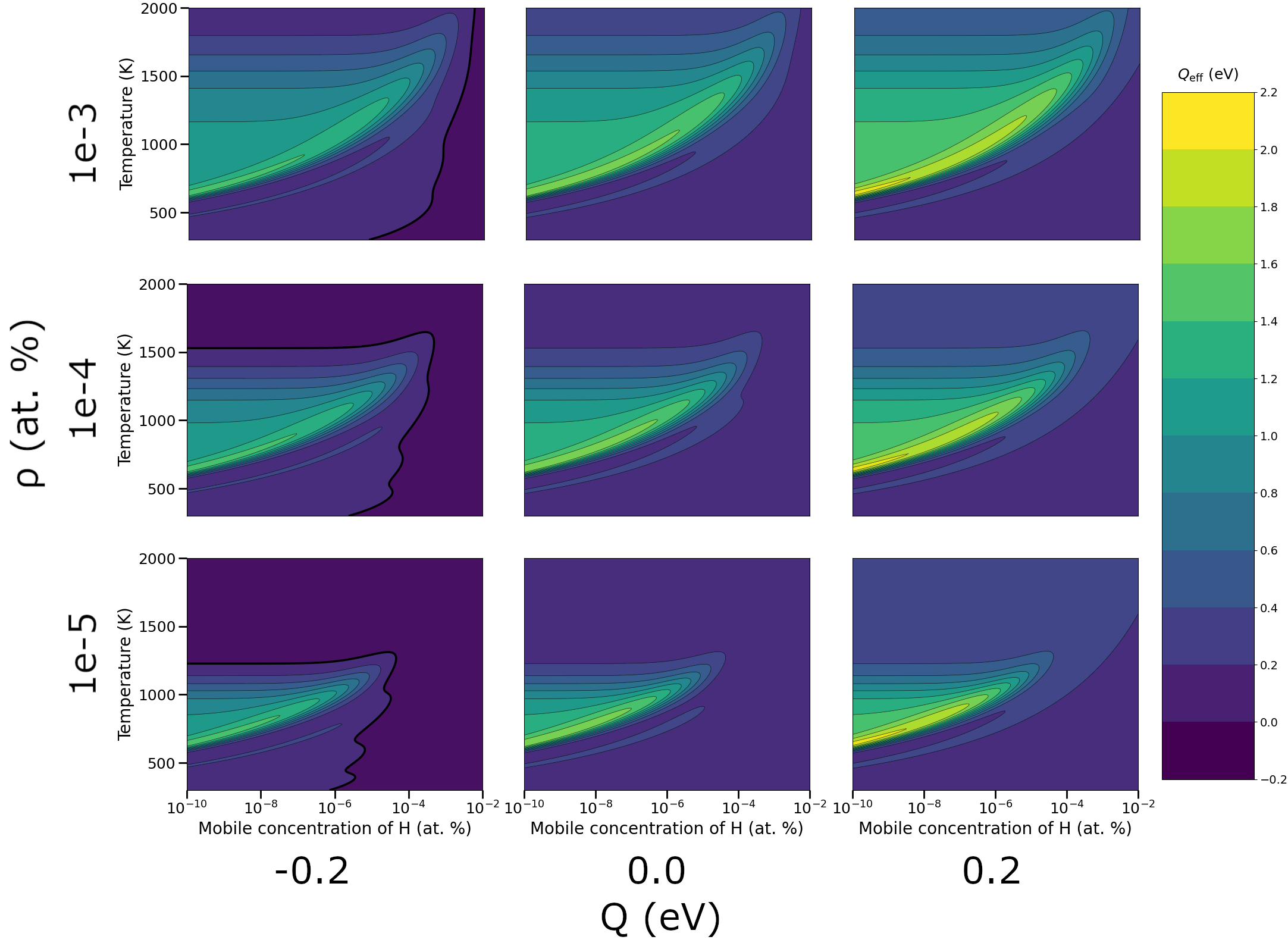}
    \caption{Indivudual panels: $Q_\mathrm{eff}$ as a function of mobile hydrogen concentration (x-axis) and temperature (y-axis) evaluated from Eq. \ref{eqn:eff-Q} using binding energies from Mason et. al. 2023 potential \cite{mason2023empirical}. True $Q$ increases from left to right (Q=-$E_m$,0 and $E_m$ where $E_m$ is migration barrier which for this potential is 0.216eV) and trap density from bottom to top (1e-5 1e-4 1e-3 at. \%) }
    \label{fig:Q_4dim}
\end{figure*}

Because the balance between trapped and mobile concentrations is temperature, hydrogen concentration, and trap concentration dependent, $Q_{\mathrm{eff}}(c_H(z),T(z),c_{\mathrm{Vac}}(z),Q)$ is a function of all these as well. Figure~\ref{fig:Q_4dim} shows $Q_{\mathrm{eff}}$, calculated using Eq.~\ref{eqn:eff-Q}, as a function of the local mobile hydrogen concentration and temperature for three vacancy concentrations and values of $Q$.

$Q_\mathrm{eff}$ is dominated by the change of trap occupancy with temperature, which is the only contribution to $Q_\mathrm{eff}$ in the middle panels of Fig. \ref{fig:Q_4dim} as $Q=0$ removes the other terms in Eq. \ref{eqn:eff-Q}. Increasing vacancy concentration broadens the $Q_\mathrm{eff}$ peaks, meaning $Q_\mathrm{eff}$ is higher in broader range of temperatures and concentrations with higher vacancy concentration. $Q_\mathrm{eff}$ approaches the true $Q$ value at high concentrations or high temperatures, where the trapped to mobile hydrogen ratio becomes small. $Q_\mathrm{eff}$ peaks at moderate/high temperatures because the average occupancy only starts to change when the temperatures are able to activate the detrapping processes. 

In summary, if traps are present, this leads to a wide range of different inferred $Q$ values depending on the exact details of experimental sample quality, temperature range, hydrogen loading etc. 

\subsubsection{MD with Vacancies Present}

\begin{figure}[H]
    \centering
    \includegraphics[width=1\linewidth]{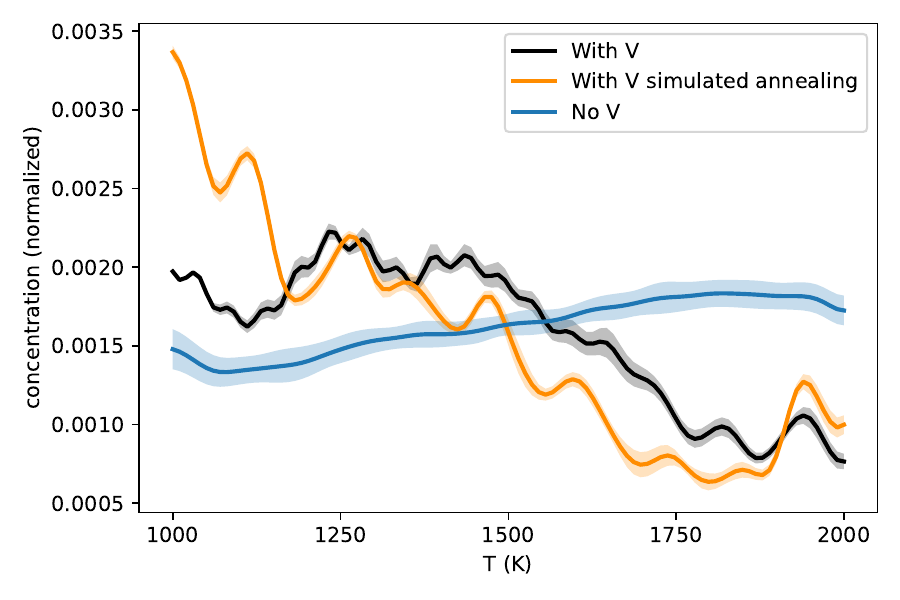}
    \caption{Normalised hydrogen concentration as a function of local temperature (1000-2000 K) obtained from molecular dynamics simulations. The curves correspond to a system containing 0.4 at.\% vacancies arranged on a regular lattice after 20 ns of evolution from an initially homogeneous hydrogen distribution (black), the same system after simulated-annealing described in the text (orange), and a vacancy-free reference system (blue). Hydrogen concentration profiles are represented using kernel density estimation (KDE), with bandwidth of 7.3 Å, which produces smoother concentration fields and reduces layer-to-layer variations associated with the ordered vacancy arrangement.}
    \label{fig:md_traps_vs_no_traps}
\end{figure}

As in KMC, similar trapping effects are present in MD simulations as well, as see in Fig.~\ref{fig:md_traps_vs_no_traps}. In contrast to the results in absence of vacancies, when vacancies are introduced in the MD simulations, the effective heat of transport $Q_{\mathrm{eff}}$ remains mostly positive, and hydrogen accumulates in the colder regions.


The presence of vacancies significantly reduces the effective diffusivity, resulting in much longer times required to reach the steady state. Consequently, vacancy-containing systems do not reach steady state within the 20 ns simulation time when starting from a homogeneous hydrogen concentration. To accelerate equilibration, a second dataset was generated using initially elevated thermostat temperatures of 3000 K (hot) and 2000 K (cold), which were gradually reduced to 2000 K and 1000 K, respectively, over 10 ns. The system was then evolved for an additional 10 ns under the final temperature gradient. The resulting vacancy-containing profiles, together with the original vacancy-free profile, are shown in Fig.~\ref{fig:md_traps_vs_no_traps}. 

\begin{figure}[H]
    \centering
    \includegraphics[width=1\linewidth]{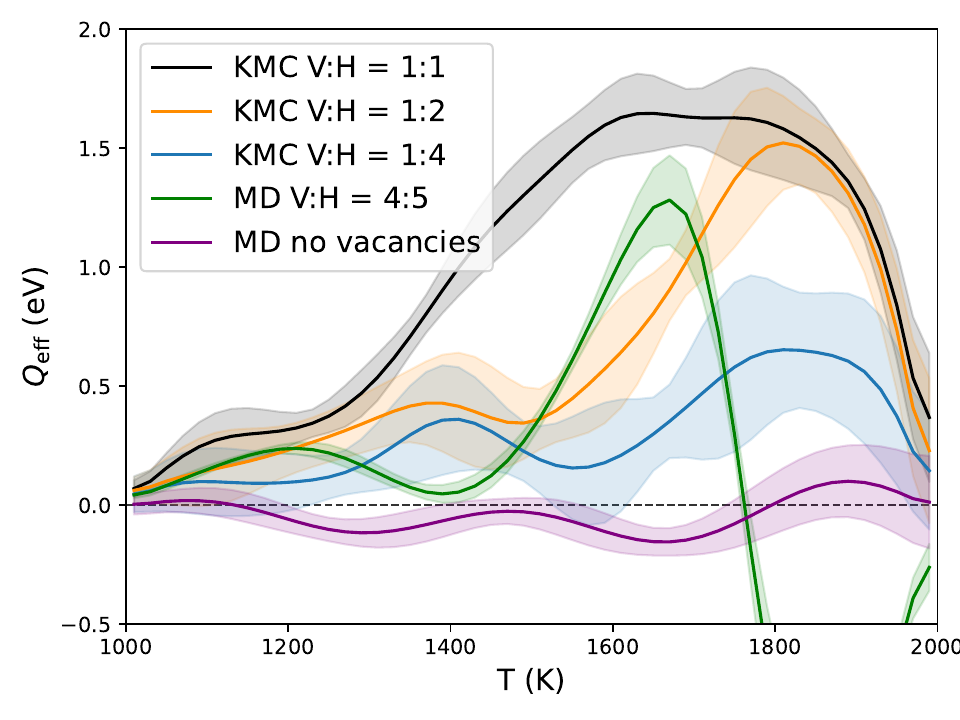}
    \caption{$Q_\mathrm{eff}$ measured directly from KMC and MD simulated concentration profiles of hydrogen using Eq. \ref{eqn:eff-Q-definition}. KDE (3.5 Å bandwidth) was used to represent concentration profiles from KMC data. MD data was coarse grained using truncated inverse Fourier transform to remove high frequency features from the regular vacancy grid placement. Shaded band in both cases represent standard deviation between runs in KMC and between time chunks of single trajectories in MD. Minimum at $\sim 1900K$ in MD with vacancies run, and boundary effects are discussed in the main text. MD data with vacancies is taken from the run with simulated annealing.}
    \label{fig:Q_compare}
\end{figure}

Fig.~\ref{fig:Q_compare} shows $Q_\mathrm{eff}$ measured from KMC and MD simulations with and without vacancies present. Both KMC and MD simulations show positive $Q_\mathrm{eff}$ values in the magnitude of 1 eV in contrast to and order of magnitude weaker negative $Q$ from MD simulations without vacancies. The trends are consistent between MD and KMC even though the sign of true $Q$ is different (slightly negative in MD and $\sim$0.2 eV in KMC). This implies that inferred $Q_\mathrm{eff}$ from total concentrations is dominated by changes in trapping rather than bias in jumps, which dominates the vacancy free diffusion in these systems.

The data for $Q_\mathrm{eff}$ in Fig. \ref{fig:Q_compare} contains two features that require further explaining: there is a fairly large negative $Q_\mathrm{eff}$ peak in MD data with vacancies, and $Q_\mathrm{eff}$ approaches zero at the edges. While vacancy mobility is low and no large voids have time to form, we nevertheless observe the formation of H clusters that are larger than would be stable in a mono-vacancy. The associated concentration peaks can be traced to clusters that remain at approximately fixed spatial locations over the duration of the simulation. We do not imply that these locations are preferred; rather, the observation serves to confirm that the peaks are not the result of data post-processing or inadequate sampling time, but correspond to a some unidentified small high temperature effect (e.g. the formation of a few divacancies). These H clusters are the source of the small concentration peak in the higher temperature region. This small peak results in large negative peak in $Q_\mathrm{eff}$ in Fig. \ref{fig:Q_compare}. While the concentration peak itself is relatively small, $Q$ is calculated from the relative slope $\nabla c/c$ which for low $c$ results in large absolute values of $Q$ even from small peaks in the data. 

Because KDE representation of concentration profiles gives smooth peaks near thermostats, and generally because of symmetry around the thermostats, $\nabla c/c \rightarrow 0$ $Q$ approaches zero in both 1000 K and 2000 K points in Fig. \ref{fig:Q_compare}. This is a property of the simulation geometry rather than that of $Q_\mathrm{eff}$. $Q_\mathrm{eff}$ can of course be 0 at these points, but this can not be inferred from the data presented here. 

\subsection{Implications for Continuum-Scale Model Parametrization}

Hydrogen permeation can be simulated in continuum models using at least two different approaches. In the first, trapping is treated explicitly by transiently evolving both the mobile and trapped hydrogen populations throughout the simulation domain. In the second, only one population, e.g. the mobile hydrogen concentration, is evolved. The trapped population is then determined from the mobile concentration by assuming rapid local equilibrium between mobile and trapped states.

The first approach can be characterized with the following set of equations:

\begin{align}
    \frac{\partial x}{\partial t} &= -\nabla \cdot J_x - r(x, c_\mathrm{trapped},\rho) \\
    \frac{\partial  c_\mathrm{trapped}}{\partial t} &= r(x, c_\mathrm{trapped},\rho)
\end{align}
where $c_\mathrm{trapped}$ is total trapped population, $x$ is mobile population and $r$ is the combined total detrapping and trapping term describing the exchange between the populations.

In the case of multi-occupancy traps or multiple trap species, $c_\mathrm{trapped}$ is further divided into multiple populations and this typically becomes an equation with a transition matrix.

The second approach is characterized by a single equation \cite{oriani1970diffusion, kaur2025effect, mason2026multiscale}:
\begin{equation}
    \left( 1 + \rho\frac{\text{Var}(s)}{x} \right)\frac{\partial x}{\partial t} = -\nabla \cdot J_x 
\end{equation}
An additional source term is needed if temperatures or trap concentration change over time as they change the balance between trapped and mobile concentrations. This approach transforms the problem from diffusion-trapping problem into a heat equation like form. Here, the factor $1 + \rho\frac{Var(s)}{x}$ plays a role analogous to a temperature-dependent volumetric heat capacity in the heat equation. Just as heat capacity determines how strongly temperature responds to a given heat flux, this factor determines how strongly the local mobile concentration responds to an incoming hydrogen flux. A larger value indicates that a greater fraction of the incoming hydrogen is absorbed into the traps, reducing the change in mobile concentration.

In both approaches 
\begin{equation}
    J_x=-D\left(\nabla x +x\frac{Q\nabla T}{k_\mathrm{B}T^2}\right)
\end{equation}
where $Q$ need to be interpreted from the slope of mobile concentration. $Q_\mathrm{eff}$ inferred from total concentrations should be avoided in both formulations. 

The total hydrogen concentration includes both mobile hydrogen and hydrogen trapped at defects. To determine the mobile fraction, one must know the concentration and free energy of each trapping site. These quantities define the equilibrium occupancy of the traps, from which the amount of trapped and mobile hydrogen can be calculated.

If estimation of $Q$ from experiments is not possible, either because the concentration of trapping sites and hence mobile hydrogen concentration is unknown or because sufficient experimental data are unavailable, we recommend using $Q=\Delta E_\mathrm{mig}$ corresponding to the dominant diffusion pathway as an order of magnitude approximation. This is consistent with a kinetic Monte Carlo model where hopping rates are based on local temperature but where diffusion barriers are constant. Alternatively, a linear temperature dependence $Q=Q_0+aT$ often assumed in the literature seems reasonable. If $a=\frac{\Delta E_\mathrm{min}}{\Delta T}$, the temperature-dependent energy change may arise, for example, from changes in elastic interaction energies associated with thermal expansion. Based on the MD results presented here, elastic energy changes are likely to underestimate the temperature dependence of $Q$. A direct fit to MD data may provide a more accurate estimate, but the simulations should be performed at the relevant temperature ranges as high and low temperature behavior may differ. Signal to noise ratio of $Q$ also becomes small if $Q$ is measured near $Q=0$ as concentration fluctuations from ordinary diffusion dominate the signal. 

These fits may not be consistent between different simulation setups and are dependent on the interatomic potential used. Though not explicitly tested here, we also expect isotope effects in the $Q$ estimates, based on observed differences between mass 10 u and 1 u hydrogen simulations. 

It should also be recognized that diffusion may be governed by pathways other than the bulk lattice, such as grain boundaries. While variations in diffusivity are not the only factor influencing $Q$, they are likely to play a major role in it. Hence estimates of $Q$ are only meaningful if the MD simulations reproduce the diffusion mechanism relevant to the material of interest, since $Q$ depends on the diffusion pathway. There is unlikely to be an universal $Q$ valid for both polycrystalline and single crystal materials if the dominant diffusion pathway changes.

Finally, none of the non-experimental approaches discussed above account for electronic effects, which may be important in some materials, but are out of the scope of the present work. 

\section{Conclusions}

Molecular dynamics simulations show that $Q$ is positive at lower temperatures. These simulations show a sign flip at $\sim1000$ K for protium, but when further investigated with artificial mass 10 u hydrogen, $Q$ stays positive for the whole simulated temperature range. Together with the observation that the temperature dependence of the elastic energy is too small to account for the variation in $Q$, this implies that the linear behavior of $Q$ goes beyond quasi-HTST effects that are not easily described analytically. Mass dependence also implies possible isotope effects.

As the measured effective heat of transport, $Q_\mathrm{eff}$, reflects the gradient of the total hydrogen concentration profile, it is largely governed by trapping when trapped hydrogen dominates the overall concentration (e.g. from strong binding to traps or from high trap density). As a result, trapping alone can generate an apparent heat-of-transport response in zero-flux conditions even without including an explicit thermodiffusion term in the mobile-hydrogen flux. 

Continuum modeling should use $Q$ determined from mobile hydrogen concentration profiles. A portion of the measured $Q_\mathrm{eff}$ may have arisen from the trapped hydrogen population and is already accounted for in models that explicitly include hydrogen trapping and detrapping. This contribution should therefore not be included a second time in the mobile flux term. 

\section*{Acknowledgments}
This work is part of the UKAEA’s Lithium Breeding Tritium Innovation (LIBRTI) programme, funded by the Department of Energy Security and Net Zero, as announced by the UK Government in October 2023. 

We thank Felix Hofmann and Daniel J. Long for a stimulating discussion.

\section*{Author contribution statement}
V.J. Conceptualization, Investigation, Formal Analysis, Validation, Software, Methodology, Visualization, Writing - Original Draft, D.R.M. Conceptualization, Formal Analysis, Validation, Software, Writing - Review \& Editing, Supervision, S.K. Formal Analysis, Writing - Review \& Editing.

\section*{Data availability}
Data will be made available upon publication.

\section*{Declaration of Generative AI}
Microsoft Copilot (GPT‑5 chat model) was used in the preparation of data-analysis scripts, which were reviewed by the authors, and to improve the language of the manuscript. The authors take full responsibility for the accuracy and correctness of the data and the text.

\appendix

\section*{Appendices}

\section{Linear $Q$ Resulting from Linear in T Change of Minima and Maxima of the Potential Energy Landscape}\label{app:linear_in_T_Q}

In 1D, (nearest neighbor) transition probabilities within a time interval can be described as a tri-diagonal matrix:

\[
\resizebox{1.0\columnwidth}{!}{$
P=\begin{pmatrix}
- p_{1,2} & p_{2,1} & 0         & 0         & 0 \\
p_{1,2}     &  - (p_{2,1}+p_{2,3}) & p_{3,2} & 0         & 0 \\
0           & p_{2,3} &  - (p_{3,2}+p_{3,4}) & p_{4,3} & 0 \\
0           & 0       & p_{3,4}   &  - (p_{4,3}+p_{4,5}) & p_{5,4} \\
0           & 0       & 0         & p_{4,5}   &  - p_{5,4}
\end{pmatrix}
$}
\]
\vspace{0.5em}

\begin{figure}[H]
    \centering
    \includegraphics[width=1\linewidth]{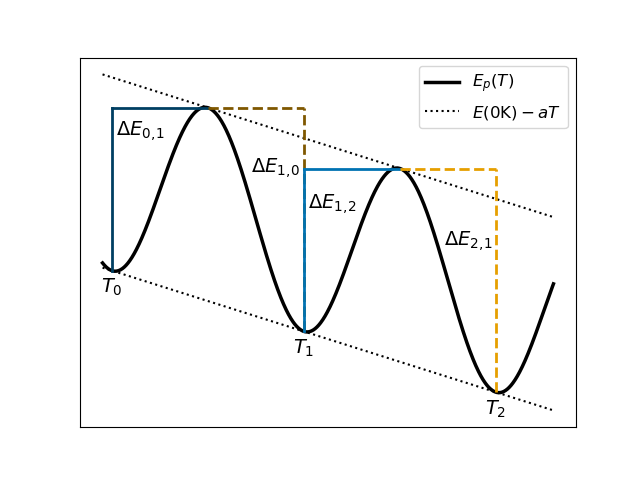}
    \caption{Schematic for asymmetric barriers. Left and right barriers stay the same. In a simplified model, asymmetry follows from linear change in potential energy as a function of temperature.}
    \label{fig:energy_landscape_scheme}
\end{figure}

If the asymmetry arises from an asymmetric potential energy landscape, such that the energy barriers for left and right transitions differ, as seen in the sloped energy landscapes observed for hydrogen in palladium in Ref.~\cite{schelling2012}, we can try to approximate corresponding $Q$. If we assume that the transition rates $p_{ij}$ follow an Arrhenius form, with potential energy minima and saddle points that vary linearly with temperature as schematised in Fig.~\ref{fig:energy_landscape_scheme},
\begin{equation}
E_{\mathrm{min/sad}}(T) = E_{\mathrm{min/sad}}(0\,\mathrm{K}) + aT.
\end{equation}
we write
\begin{align}
    \frac{1}{D}\frac{\partial D}{\partial T} & = \notag \\ 
    \frac{\partial \log(D)}{\partial T} &= \frac{\log(p_{n+1,n}) - \log(p_{n,n+1})}{T_{n+1}-T_{n}},\\
    p_{n,n+1} &= v_0\exp\!\left(\frac{-\Delta E - 0.5a(T_{n+1}-T_n)}{k_{\mathrm{B}}T_{n}}\right),\\
    p_{n+1,n} &= v_0\exp\!\left(\frac{-\Delta E + 0.5a(T_{n+1}-T_n)}{k_{\mathrm{B}}T_{n+1}}\right),
\end{align}
and therefore
\begin{equation}
    \frac{\partial \log(D)}{\partial T}
    = \frac{0.5a(T_{n+1}+T_n)+\Delta E}{k_{\mathrm{B}}T_{n+1}T_n}.
\end{equation}

\begin{figure}[H]
    \centering
    \includegraphics[width=1\linewidth]{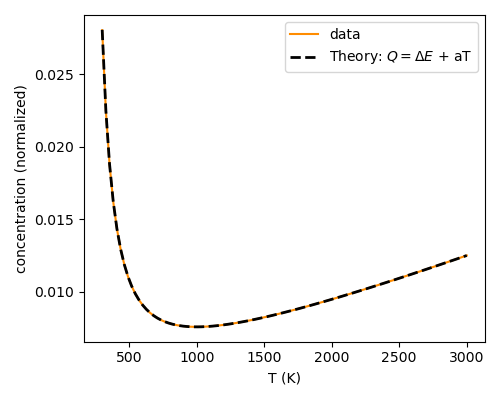}
    \caption{Analytic solution (dashed black) for hydrogen concentration when $Q=\Delta E + aT$ compared with steady state solution of the corresponding tridiagonal transition matrix (orange).}
    \label{fig:tridiagonal_matrix_vs_analytic}
\end{figure}

It is not immediately obvious that we can omit the bias term from Eq. \ref{eq:Q_general_diffusivity} and infer the diffusivity from the difference between the left- and right-going hopping rates. 
Strictly speaking, a single diffusivity does not meaningfully exist in this context, and the fundamental description of the system is given by the hopping rates in the transition matrix. Thus, the above argument is intended to provide intuition rather than a rigorous derivation. However, it accurately reproduces the steady-state solution of the corresponding transition matrix (Fig. \ref{fig:tridiagonal_matrix_vs_analytic}) and correctly predicts the temperature dependence of $Q$ in the steady state of that transition-matrix.

If we write $T_n=T$ and $T_{n+1} = T + \Delta T$. Then
\begin{equation}
    k_\mathrm{B}T^2\frac{\partial \log(D)}{\partial T}
    = (\Delta E + aT_n + 0.5a\Delta T)\gamma,
\end{equation}
where if $\Delta T<<T$, $\gamma=(1+\Delta T/T)^{-1} \sim 1$ and if $|0.5a\Delta T|<< |\Delta E|$ then
\begin{equation}
    Q \simeq \Delta E + aT.
\end{equation}

This behaviour can be understood by comparing the competing transition rates, $v_0 \exp(-E_1 / k_{\mathrm{B}} T_1)$ and $v_0 \exp(-E_2 / k_{\mathrm{B}} T_2)$. The dominant contribution arises from the ratios $E/T$, i.e. the exponents $-E_1 / (k_{\mathrm{B}} T_1)$ and $-E_2 / (k_{\mathrm{B}} T_2)$. The difference is controlled by both the energy and temperature differences.

At low temperatures, a fixed temperature difference $\Delta T$ produces a comparatively large relative change in $T$, and thus strongly affects the ratios $E/T$. At higher temperatures, the same $\Delta T$ leads to a much smaller relative change, and the constant energy difference between the barriers becomes the dominant factor. This competition leads to a linear temperature dependence of $Q$.

We did not implement bias to diffusion barriers in our KMC code. Instead, in Fig. \ref{fig:tridiagonal_matrix_vs_analytic} we compare the numerical steady state solution of our tridiagonal diffusion matrix and compare it to the analytical expression in Eq. \ref{eq:non_symmetric_solution} and we see a perfect match. Tridiagonal matrix was constructed with $\Delta E=$ 0.1 eV and $a=-$1e-4 eV where transition rates are $p_{n,n+1} = v_0\exp\!\left(\frac{-\Delta E - 0.5a(T_{n+1}-T_n)}{k_{\mathrm{B}}T_{n}}\right)$ and $p_{n+1,n} = v_0\exp\!\left(\frac{-\Delta E + 0.5a(T_{n+1}-T_n)}{k_{\mathrm{B}}T_{n+1}}\right)$.

\section{Derivation of $Q_{\mathrm{eff}}$}\label{app:qeff_derivation}

The average occupancy of a trap at point $\textbf{r}$ and time $t$ is the expectation $\langle s(\textbf{r}, t) \rangle = \sum_{i=0}^n i y_i(\textbf{r}, t)$ across occupancies $i=0,\dots, n$. The probability distribution is defined by the mobile concentration and temperature fields $y_i(\textbf{r}, t) = y_i(x(\textbf{r}, t), T(\textbf{r}, t))$. The trapped concentration field is therefore $c_t(\textbf{r}, t) = \rho(\textbf{r})\langle s(\textbf{r}, t) \rangle$. \\

The trap relaxation time is defined by the trapping and detrapping rates. If the trap relaxation time for a given $x$, $T$ is much shorter than the rate of change of $x$, $T$, adiabatic elimination can be used: the trap occupancy probability follows the equilibrium manifold $y^{\text{eq}}_i(x(\textbf{r}, t), T(\textbf{r}, t))$ defined by mobile concentration and temperature as these variables evolve.

For a steady state probability distribution $y^{\text{eq}}_i(x(\textbf{r}, t), T(\textbf{r}, t))$, taking the derivative of expected trap occupancy with respect to $x$, $T$ produces the fluctuation-response relations:

\begin{align}
    \label{eqn:fluctuations}
    \frac{d \langle s \rangle}{dx} &= \frac{\text{Var}(s)}{x} \\
    \frac{d \langle s \rangle}{dT} &= - \frac{\text{Covar}(s, \epsilon)}{k_B T^2}
\end{align}

where $\text{Var}(s) = \langle s^2\rangle - \langle s\rangle^2$ and $\text{Covar}(s, \epsilon) = \langle s \epsilon \rangle - \langle s\rangle \langle \epsilon \rangle$. These measure the susceptibility of trap occupancy to changes in its thermodynamic driving variables. The response to mobile concentration (or chemical potential) is governed by variance of occupancy, while the response to temperature is governed by the covariance between cumulative binding energy and occupancy.

For a mobile concentration field defined by temperature itself $x(T)$ and a temperature field not evolving in time, for constant $Q$, we can write the temperature derivative in equation \ref{eqn:total_c_derivative} as

\begin{align}
    \frac{\partial c}{\partial T} &=  \frac{\partial x}{\partial T} + \rho \frac{d}{dT} \langle s(x(T), T) \rangle \\
    & = \frac{\partial x}{\partial T} + \rho \left[  \frac{\partial \langle s \rangle}{\partial x} \frac{\partial x}{\partial T} + \frac{\partial \langle s \rangle}{\partial T} \right] \\
    & = \left[ 1 + \rho \frac{\text{Var}(s)}{x} \right] \frac{\partial x}{\partial T} - \rho \frac{\text{Covar}(s, \epsilon)}{k_B T^2} \\
    &= - \frac{Q}{k_B T^2} \left[  x + \rho \text{Var}(s) \right] - \rho \frac{\text{Covar}(s, \epsilon)}{k_B T^2}
\end{align}

Substituting into the definition of effective heat of transport equation \ref{eqn:eff-Q-definition} produces equation \ref{eqn:eff-Q}. The first term encodes how temperature changes $x$, which in turn changes the probability distribution. The second term encodes how temperature directly changes this distribution.

\begin{figure}[H]
    \centering
    \includegraphics[width=1\linewidth]{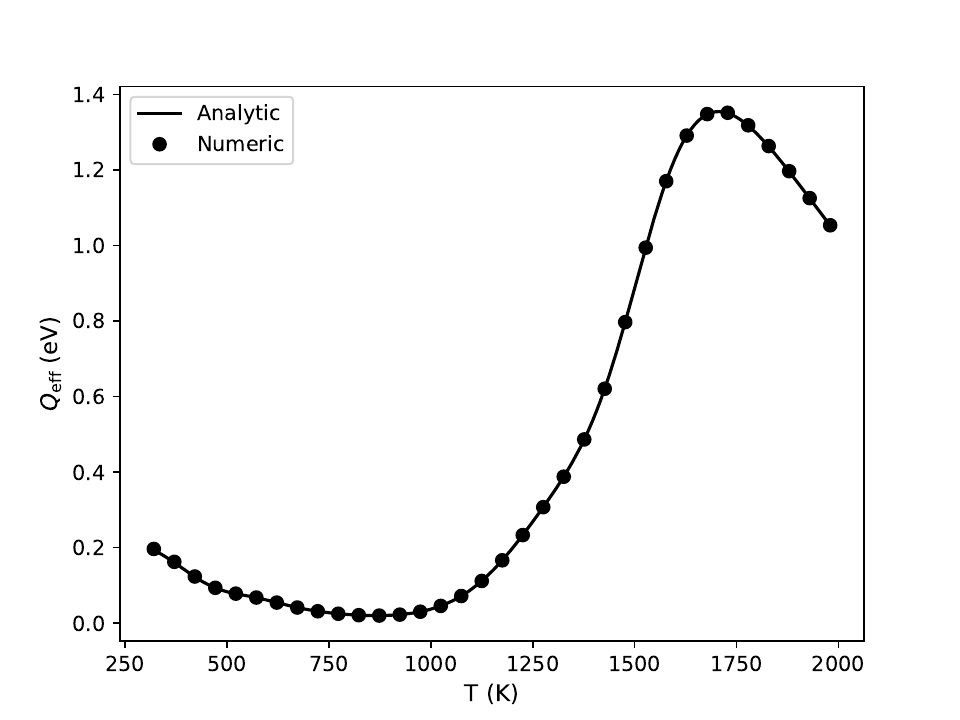}
    \caption{$Q_\mathrm{eff}$ as a function of temperature from Eq. \ref{eqn:eff-Q} (solid line) and evaluated numerically showing consistency between Eq. \ref{eqn:eff-Q} and the analytic solution of the concentration profile. }
    \label{fig:sanjeet_qeff_analytic_vs_numeric}
\end{figure}

To demonstrate that this is indeed the solution for $Q_\mathrm{eff}$ we have independently solved the concentration profile, as explained in the main text, and taken $Q_\mathrm{eff}$ from direct numerical derivative to this concentration profile. The results compared to our derived equation are shown in Fig. \ref{fig:sanjeet_qeff_analytic_vs_numeric}.

\section{On Local Equilibrium Assumption}\label{app:break_of_local_eq}
It is possible that local thermodynamic equilibrium breaks down, which consequently affects HTST. In particular, one cannot self-consistently assume both: a barrier asymmetry arising from a temperature difference between diffusion barriers on either side, and a well-defined single temperature across the entire potential well. Diffusion rates are related to entropy, which in turn depends on the Boltzmann factors of different phase space coordinates within the potential well. Evaluating a Boltzmann factor at an temperature that is off by $\Delta T$ corresponds to an effective misestimation of the state energy $E$ by a factor $\frac{1}{1+\Delta T/T}$:
\begin{equation}
\exp\!\left(-\frac{E}{k_\mathrm{B}(T+\Delta T)}\right)
=
\exp\!\left(-\frac{E}{k_\mathrm{B}T}\cdot\frac{1}{1+\Delta T/T}\right).
\end{equation}

Within a single potential well corresponding to an interstitial site, $\Delta T$ can be assumed small. To first order, the resulting error scales linearly with $\Delta T/T$ while the energy difference arising from thermal expansion is constant if $\Delta T$ does not change, implying that at high temperatures local equilibrium may remain consistent with asymmetric barriers. It should also be noted that $\Delta T$ is the upper limit for temperature misestimation within one potential well. 

However, it remains unclear whether the HTST prefactor can be defined consistently under a thermal gradient. While the interstitial energy may be approximated using a local temperature because it is dominated by interactions with nearby atoms, over which the temperature variation is small, the eigenfrequencies entering the prefactor depend on collective vibrational modes extending over a much larger region. 

In addition, the phonon populations associated with these modes may not correspond to thermal equilibrium, as heat is continuously transported through the system by phonons emitted and absorbed at the thermostatted boundaries. As a result, the assumption of a single local temperature becomes less well defined and the thermodynamic quantities entering HTST lose their usual meaning. A more rigorous treatment would require a non-equilibrium framework, for example based on Boltzmann transport theory, which is beyond the scope of the present work. 







\bibliographystyle{unsrt}
\bibliography{refs}

\end{document}